# Unique prospects for graphene-based THz modulators

*Berardi Sensale-Rodriguez[a], Tian Fang, Rusen Yan, Michelle M. Kelly, Debdeep Jena, Lei Liu[b], and Huili (Grace) Xing[c]*.

Department of Electrical Engineering, University of Notre Dame, Notre Dame, Indiana 46556 USA.

E-mail address: [a] bsensale@nd.edu, [b] liu_lei@nd.edu, [c] hxing@nd.edu.

ABSTRACT    The modulation depth of 2-D electron gas (2DEG) based THz modulators using AlGaAs/GaAs heterostructures with metal gates is inherently limited to < 30%. The metal gate not only attenuates the THz signal (> 90%) but also severely degrades the modulation depth. The metal losses can be significantly reduced with an alternative material with tunable conductivity. Graphene presents a unique solution to this problem due to its symmetric band structure and extraordinarily high mobility of holes that is comparable to electron mobility in conventional semiconductors. The hole conductivity in graphene can be electrostatically tuned in the graphene-2DEG parallel capacitor configuration, thus more efficiently tuning the THz transmission. In this work, we show that it is possible to achieve a modulation depth of > 90% while simultaneously minimizing signal attenuation to < 5% by tuning the Fermi level at the Dirac point in graphene.

MANUSCRIPT TEXT

The terahertz (THz) electromagnetic spectrum has long been recognized as an important region for scientific research. However, due to the lack of devices, circuits and systems for effective THz signal generation, detection, and modulation, this region remains the least explored and developed in the entire electromagnetic spectrum. The past decade witnessed a substantial increase in THz research activities[1-2]. Electrically tunable THz modulation is one of the actively pursued subjects due to its importance in applications such as communications, imaging, and spectroscopy. Kleine et al.[3-5] demonstrated a THz modulator operating at room temperature (RT) by employing a semiconductor 2-dimensional electron gas (2DEG) structure. Though RT operation is a significant advance compared to typical devices that operate at cryogenic temperatures, poor modulation depths of 3-4 % have been reported so far. Consequently, the principal direction of the RT modulator research turned towards metamaterial-based approaches[6-8], with up to 52% modulation recently reported[7]. However, these metamaterial-based devices have several disadvantages in comparison to the 2DEG-based modulators. For example, they are intrinsically narrowband, and usually have a polarization-dependent response[8]. There has been little discussion in the literature on the fundamental limits of the performance of 2DEG-based THz modulators, and whether performance superior to that of metamaterial-based devices is theoretically feasible. In this work we show that the modulation depth of the previously proposed AlGaAs/GaAs 2DEG THz modulators is inherently limited to be < 30% due to the adverse effect of the metal gate; however, by employing graphene in place of the metal gate, modulation depth > 90% is achievable. The unique combination of electronic and optical properties of graphene works in favor, and enables the potential advance in THz technology.

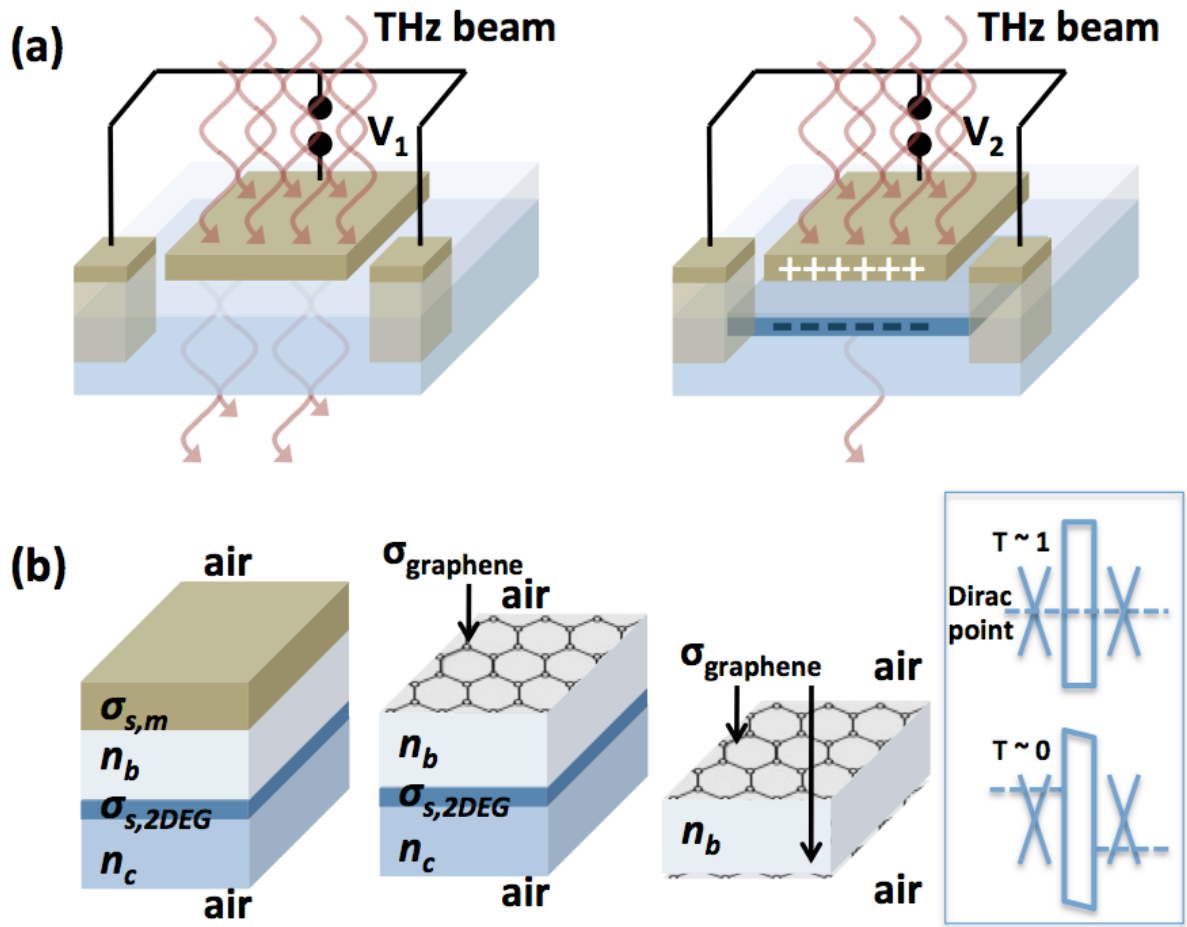

**Figure 1.** (a) Operating principle of a 2DEG-based THz modulator. THz transmission through a conducting media (2DEG) is tuned by a voltage applied between the top gate and the 2DEG. THz transmission is high with low 2DEG densities, and low with high 2DEG densities due to enhanced absorption and reflection. (b) Layer structures of traditional metal-gate/2DEG and proposed graphene/2DEG and graphene/graphene THz modulators. Shown in the box are the schematic energy band diagrams of a graphene/insulator/graphene modulator that promises near zero beam attenuation and unity modulation depth. When the Fermi level is at the Dirac point of both the top and bottom graphene layers, THz transmission approaches unity; when electron and hole sheets of charges are formed in the top and bottom graphene layers, THz transmission nears zero.

Graphene, a single layer of carbon atom with honeycomb structure, has attracted intense attention in the fields of physical, chemical and biological sciences since its discovery in 2004[9-11]. Due to its unique conical and symmetric band structure, graphene has been proposed for a rich array of novel devices and applications. Its optical properties have also been extensively explored, however, to date there are only a few studies of graphene-based THz devices in the literature including emitters[12-15], detectors[16-19] and nano-antennas[20]. In this work, we present a proposal of graphene-based THz modulators for the first time, to the best of our knowledge.

The structure of a generic 2DEG-based electrically driven THz modulator is shown in Fig.1(a). Since THz transmission through a conducting media is a function of its conductivity, modulation of THz transmission can be achieved by electrically tuning the 2DEG density using a gate. Even though side-gates have been employed in nanoscale electronic devices potentially removed from the THz beam path, a parallel capacitor-like gate is unavoidable in these 2DEG THz modulators due to their large size – at least comparable to the THz beam wavelength (> 0.1 mm). In traditional 2DEG THz modulators a metal gate is used. In this work we investigate the performance benefits of replacing the metal gate by a single-layer of graphene, as shown in Fig. 1(b). We also consider the use of two sheets of graphene separated by an insulator, taking advantage of the facile fabrication and low cost of large area graphene grown using chemical vapor deposition. The energy band diagrams of such a graphene/insulator/graphene THz modulator are also shown in Fig. 1(b). When the Fermi level is at the Dirac point of both the top and bottom graphene layers, THz transmission approaches unity (we show later that the minimal conductivity of graphene introduces negligible beam attenuation); when electron and hole sheets of charges are formed in the top and bottom graphene layers under appropriate bias conditions in the device, THz transmission approaches zero.

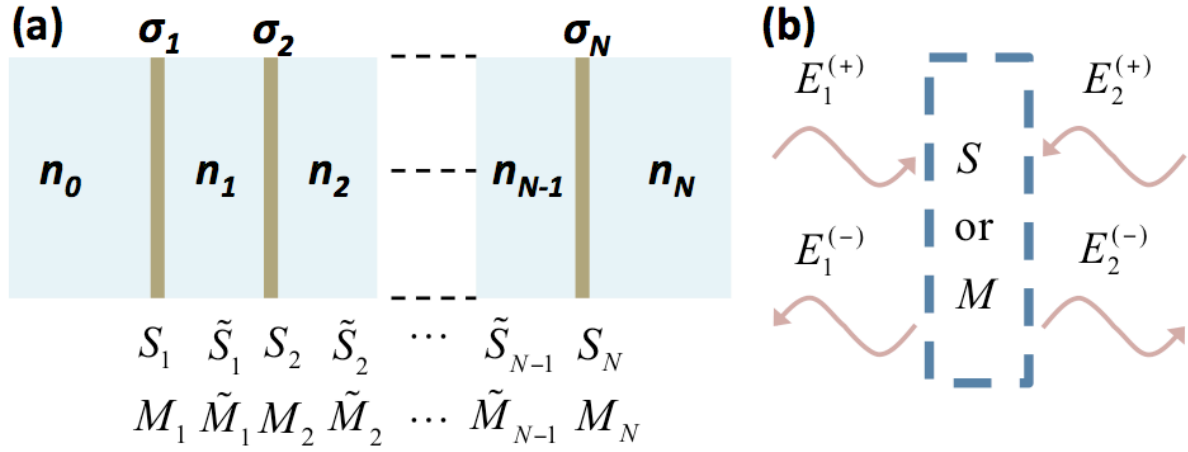

**Figure 2.** (a) Definition of parameters in a generic structure with N conductive sheets separated by dielectric interlayers for calculating THz transmission using the matrix-transfer method. The transmission through each interface and dielectric interlayer is represented by an *S* or *M* matrix. (b) Definition of the electric field vectors of incident and reflected waves at two ports of a generic optical system.

Figure 2(a) shows the schematic of a general structure with N conductive sheets with a sheet conductivity $\sigma_i$ separated by dielectric materials with optical refractive index $n_i$. To model THz transmission through this structure, the wave-transfer matrix theory is employed[21]. Each conductive interface is represented by an $S_i$ or $M_i$ matrix and each dielectric interlayer is represented by an $\tilde{S}_i$ or $\tilde{M}_i$ matrix. Shown below in Eq. (1) are the definitions of the *S*- and *M*-matrix and their relationship to the electric field vector of the incident (+) and reflected (-) waves at the two ports of an optical system as sketched in Fig. 2(b). To compute the *S*-matrix of the composite optical system shown in Fig. 2(a), the total *M*-matrix can be obtained first by multiplying the *M* matrices associated with each interface and dielectric interlayer and then converted to the *S*-matrix, as shown in Eq. (1):

$$\begin{pmatrix} E_2^{(+)} \\ E_1^{(-)} \end{pmatrix} = \begin{pmatrix} t_{12} & r_{21} \\ r_{12} & t_{21} \end{pmatrix} \begin{pmatrix} E_1^{(+)} \\ E_2^{(-)} \end{pmatrix} = S \begin{pmatrix} E_1^{(+)} \\ E_2^{(-)} \end{pmatrix}$$

$$\begin{pmatrix} E_2^{(+)} \\ E_2^{(-)} \end{pmatrix} = \frac{1}{t_{21}} \begin{pmatrix} t_{12}t_{21} - r_{12}r_{21} & r_{21} \\ -r_{12} & 1 \end{pmatrix} \begin{pmatrix} E_1^{(+)} \\ E_1^{(-)} \end{pmatrix} = M \begin{pmatrix} E_1^{(+)} \\ E_1^{(-)} \end{pmatrix}. \tag{1}$$

$$M = M_N \cdot \tilde{M}_{N-1} \cdot \ldots \cdot \tilde{M}_2 \cdot M_2 \cdot \tilde{M}_1 \cdot M_1$$

Since the metal gate thickness (~ 5 nm) and the effective thickness of the 2DEG (on the order of nanometers) are several orders of magnitude smaller than the THz beam wavelength (e.g. 500 um at 600 GHz), both the metal gate and 2DEG can be treated as zero thickness conductive sheets. In this analysis, the metal gate is modeled with a constant sheet conductivity given by $\sigma_{s,m} = \sigma_m \times t_m$, where $\sigma_m$ is the metal bulk conductivity and $t_m$ the thickness; the 2DEG is represented as a variable conductive layer with an associated sheet conductivity $\sigma_{s,2DEG}$. Additionally it is assumed that $\omega\tau \ll 1$ for all conductive sheets, where $\tau$ is the electron momentum relaxation time and $\omega$ is the angular frequency of the THz beam, so that the optical sheet conductivity equals the DC electrical conductivity: $\sigma_s(\omega) = \sigma_{s,DC}$. Under normal incidence, the Fresnel coefficients in the S-matrix for a zero thickness conductive layer located between two dielectric materials with a refractive index of $n_1$ and $n_2$, respectively, are given by[22]:

$$t_{12} = \frac{2n_1}{n_1 + n_2 + Z_0\sigma_s}, \quad r_{12} = \frac{n_1 - n_2 - Z_0\sigma_s}{n_1 + n_2 + Z_0\sigma_s}, \quad t_{21} = \frac{2n_2}{n_1 + n_2 + Z_0\sigma_s}, \quad r_{21} = \frac{n_2 - n_1 - Z_0\sigma_s}{n_1 + n_2 + Z_0\sigma_s}, \tag{2}$$

where $\sigma_s$ is the sheet conductivity of the zero thickness conductive layer and $Z_0 = 377$ Ω is the wave impedance of vacuum. For the generic structure shown in Fig. 2(a), one can see that the dielectric interlayers between the $N$ conductive sheets do not introduce electromagnetic beam attenuation though they can induce Fabry-Perot cavity type oscillatory behavior as a function of dielectric layer thickness and beam frequency. The electromagnetic wave transmission property intrinsic to the conductive sheets can be obtained independent of the cavity effect, for instance, in a structure where the sum of all dielectric interlayer thicknesses is much smaller than the beam wavelength. In this case, one can ignore the phase change of the wave in the dielectric interlayers, i.e. the $\tilde{S}$ and $\tilde{M}$ matrices. The beam transmission through the generic structure can thus be derived using the wave-transfer matrix analysis as:

$$T = \left|t_{0N}\right|^2 \cdot \frac{n_N}{n_0} = \left[\frac{2\sqrt{n_0 n_N}}{n_0 + n_N + Z_0 \sum_{i=1}^{N} \sigma_i}\right]^2. \tag{3}$$

For the 2DEG THz modulator structures considered in Fig. 1(b), the beam transmission is given by:

$$T_{metal-gate} = \left[\frac{2}{2 + Z_0 \sigma_{s,metal} + Z_0 \sigma_{s,2DEG}}\right]^2, \quad T_{graphene-gate} = \left[\frac{2}{2 + Z_0 \sigma_{s,graphene} + Z_0 \sigma_{s,2DEG}}\right]^2. \tag{4}$$

The modulation depth (MD) of the modulators, defined as: $\left(T_{\sigma=0} - T_\sigma\right)/T_{\sigma=0}$, is therefore found to be:

$$MD_{metal-gate} = 1 - \left[1 + \frac{Z_0 \sigma_{s,2DEG}}{\left(2 + Z_0 \sigma_{s,m}\right)}\right]^{-2}, \quad MD_{graphene-gate} = 1 - \left[1 + \frac{Z_0 \sigma_{graphene}}{2}\left(1 + \frac{\mu_{2DEG}}{\mu_{graphene}}\right)\right]^{-2}. \tag{5}$$

It has been assumed that the graphene conductivity can be tuned to be ~ 0 to simplify the analytical expressions since the minimum conductivity of graphene on the order of $4e^2/h$ introduces a beam attenuation of < 5% using Eq. (4). Two key observations are evident from inspection of Eq. (5): 1) A metal gate with sheet conductivity generally greater than that of the 2DEG adversely lowers the modulation depth in addition to introducing significant beam attenuation. 2) The maximum modulation depth is limited by the maximal tunable 2DEG and graphene sheet conductivity. For example, in AlN/GaN high electron mobility transistor (HEMT) structures the maximum 2DEG sheet conductivity ($\sigma_{s,2DEG,max}$) reported in the literature is on the order of 6 mS for a 2DEG concentration of $2\times10^{13}$ cm$^{-2}$ and mobility of 2000 cm$^2$/Vs (see Table 1), thus, $Z_0 \sigma_{s,2DEG,max} \sim 2.3$.

Also assumed in Eq. (5) is that the conductivities of the graphene gate and the 2DEG channel are tuned to their minimum or maximum values simultaneously in order to achieve the maximal modulation depth. To meet this requirement, the graphene minimum conductivity (when the Fermi level is at the Dirac point) should be achieved at the threshold or subthreshold bias region of the HEMT. Above the threshold condition, the high mobility 2DEG is induced in the semiconductor channel and the opposite charge with the same concentration, a 2D hole gas (2DHG), is induced in graphene. The simultaneous tuning requirement makes graphene the best candidate for the tunable conductive pair in 2DEG-based THz modulators as the hole mobility in graphene is comparable to the electron mobility due to its symmetric and conical band structure. Furthermore, the graphene electron mobility is among the highest for a given 2DEG concentration in all 2DEG systems including InAs and InSb based heterostructures, while the hole mobility in conventional semiconductors is generally orders of magnitude smaller than that in graphene. Likewise, a graphene-graphene pair can be used for 2DEG THz modulators as shown in Fig. 1(b). It is also worth noting that optical absorption in graphene in the THz range is dominated by intraband transitions instead of interband transitions, and its optical conductivity closely follows the electrical conductivity[23-24]. Therefore graphene can be treated as a conductive sheet for THz modulation.

Based on Eq. (5) the attainable modulation depth as a function of sheet conductivity is presented in Fig. 3(a) for metal-gate/2DEG and graphene/2DEG modulators. Various material parameters used in the calculation are listed in Table 1. To facilitate the evaluation, mobility and conductivity values are compared at two carrier concentrations: 1) $5 \times 10^{12}$ cm$^{-2}$ since the highest conductivity reported in AlGaAs/(In)GaAs and graphene 2DEGs is in the neighborhood of this concentration, and 2) $2 \times 10^{13}$ cm$^{-2}$ since it represents a typical value for the highest carrier concentration achievable in Al(InGa)N/GaN, SiO$_2$/Si and graphene 2DEGs. Due to the symmetric band structure of graphene, the reported mobility values are used for both electron and holes in graphene.

Table 1. Material systems compared in Fig. 3 and their 2DEG properties at two carrier concentrations estimated from the literature

| | $n_s = 5 \times 10^{12}$ cm$^{-2}$ | | $n_s = 2 \times 10^{13}$ cm$^{-2}$ | | $\mu_{2DEG} / \mu_{graphene}$ ($5 \times 10^{12}$ cm$^{-2}$) | $\sigma_{2DEG,max} + \sigma_{graphene,max}$ |
|---|---|---|---|---|---|---|
| | $\mu$ (cm$^2$/V.s) | $\sigma$ (mS) | $\mu$ (cm$^2$/V.s) | $\sigma$ (mS) | | |
| Al(InGa)N/GaN[25-26] | 2,000 | 1.6 | 2,000 | 6.4 | 2/25 | 26.4 |
| SiO$_2$/Si[27] | 200 | 0.16 | 100 | 0.32 | 0.2/25 | 20.16 |
| AlGaAs/(In)GaAs[28] | 5,000 | 4 | - | - | 5/25 | 24 |
| Graphene[29-30] | 25,000 | 20 | 2,500 | 8 | - | 40 |

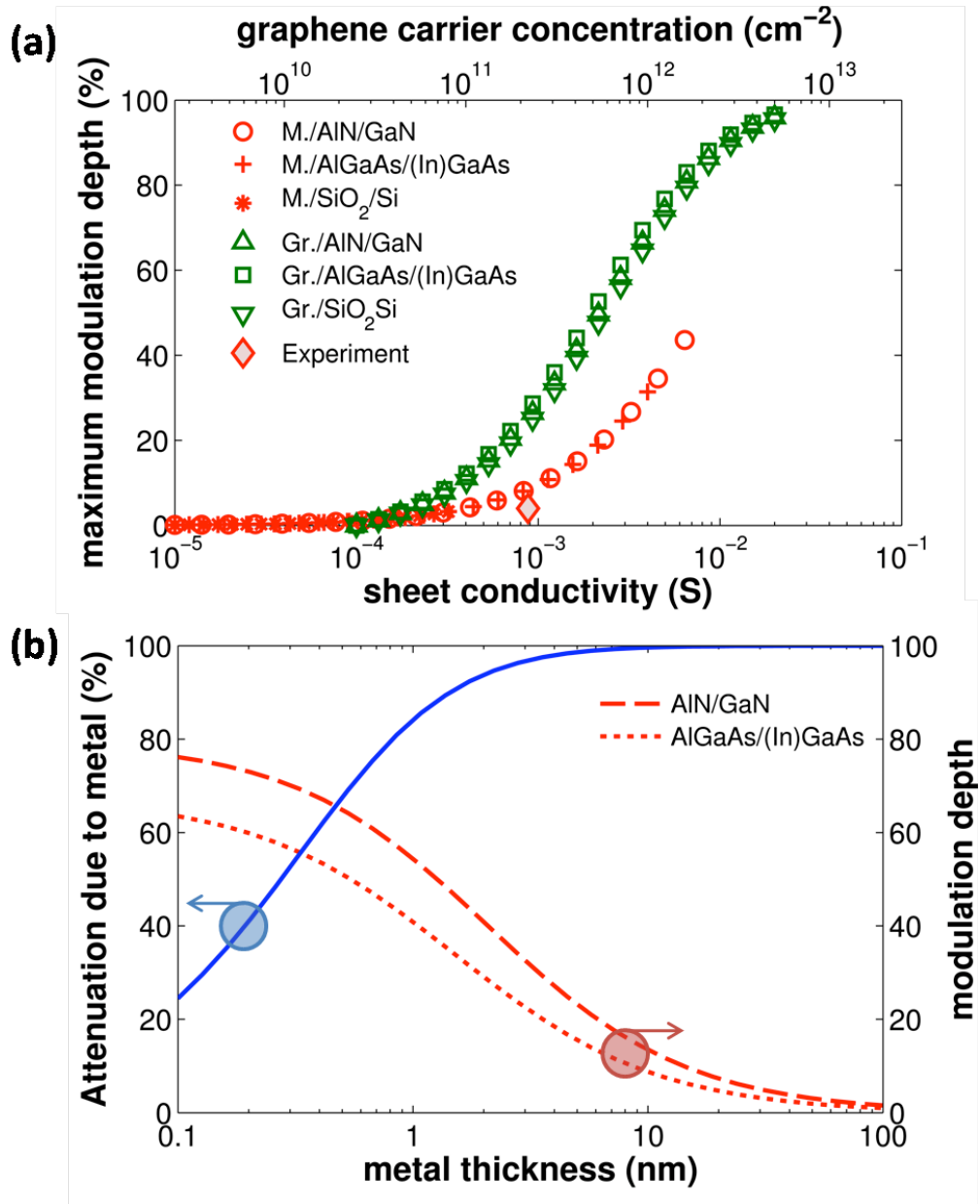

**Figure 3.** (a) Calculated achievable modulation depth for several modulator structures. For the metal/2DEG modulators, the 2DEG sheet conductivity is varied and a beam attenuation (in terms of power) of 90% due to metal is assumed; for the graphene/2DEG modulators, the graphene sheet conductivity is varied by assuming a constant $\mu_{2DEG}/\mu_{graphene}$ (at $n_s = 5 \times 10^{12}$ cm$^{-2}$) as listed in Table 1. (b) Calculated beam power attenuation introduced by the metal and maximum modulation depth for metal/AlN/GaN and metal/AlGaAs/GaAs modulators as a function of Cr gate thickness assuming a Cr bulk conductivity of $8 \times 10^6$ S/m and $\sigma_{2DEG,max}$ values listed in Table 1.

For the metal/2DEG modulators, a beam power attenuation of 90% due to the metal gate is assumed because our analysis of the experimental results reported by Kleine[3-5] et al shows that the THz beam attenuation due to the Cr-gate alone is close to 90%. Using the maximum values of the 2DEG conductivity, modulation depths of up to 5%, 30% and 45% are expected in $SiO_2$/Si, AlGaAs/(In)GaAs and Al(InGa)N/GaN based metal/2DEG modulators, respectively. It is evident that the higher the maximum 2DEG conductivity, the higher the achievable modulation depth. Also shown in Fig. 3(a) is the modulation depth observed in Kleine's experiment[3-5], which is averaged over a broad THz spectrum in the time domain approach employed and is thus subject to the cavity effect. The experimental value is lower than the calculated modulation depth since the cavity effect has been neglected in our models; hence, our calculated results represent the maximal achievable modulation depth in optimized device geometries.

For the graphene/2DEG modulators, a constant ratio of $\mu_{2DEG}/\mu_{graphene}$ at $5\times10^{12}$ cm$^{-2}$ is used as a parameter (Table 1) to simplify the comparison while varying the sheet conductivity of graphene, i.e. its carrier concentration. From Eq. (5) we see that the ratio of $\mu_{2DEG}/\mu_{graphene}$ qualitatively represents the contribution of graphene to THz modulation relative to that of the semiconductor 2DEG. Though $\mu_{2DEG}/\mu_{graphene}$ varies with carrier concentration and consequently the relative contributions of graphene and 2DEG vary, the maximum achievable modulation depth is determined by the maximum sheet conductivity sum of graphene and 2DEG. Due to the extraordinarily high sheet conductivity obtainable in graphene, the contribution of graphene in THz modulation is dominant, e.g. in a graphene/Si-2DEG structure, 99% of the resultant THz modulation can be attributed to tuning the graphene conductivity.

To further illustrate the adverse effect of the metal gate or an element with non-tunable beam attenuation in the device (e.g. cavity effect), we show in Fig. 3(b) the calculated beam attenuation and the resultant maximum achievable modulation depth as a function of the chromium gate thickness using a Cr bulk conductivity of $8 \times 10^6$ S/m. Beam attenuation quickly increases from ~ 30% to ~ 100% when the Cr thickness is increased from 0.1 nm to 10 nm; consequently, the modulation depth by tuning the 2DEG conductivity drops from near 70% to 10%. The modulation depth saturates as the beam attenuation decreases to values smaller than 10%, which again confirms that it is reasonable for us to neglect the effect of minimum conductivity of graphene in the analysis presented in this work. In practice, it is difficult to deposit continuous and uniform metal thin films of nanometers thick; moreover, thin film conductivity generally decreases with film thickness. This most likely explains the beam attenuation of 90% by a 5-nm thick Cr gate observed in Kleine's experiment[3-5]. It is challenging to reduce the beam attenuation to be lower than 90% using the conventional metal gate, which is in a stark contrast to a beam attenuation of lower than 5% introduced by the typical minimum conductivity in graphene.

In conclusion, we have presented an analytical study on the current limits of performance of 2DEG-based THz modulators, and how incorporating graphene as the 'tunable metal' gate holds promise for significant improvements in the performance. In the previously proposed metal/AlGaAs/2DEG/GaAs structures, the maximum modulation depth is inherently limited to be < 30% by the adverse effect of the highly conductive gate metal as well as the maximum achievable 2DEG sheet conductivity. A single layer graphene can be nearly transparent when its Fermi level is tuned at the Dirac point and block almost 100% THz beam when tuning to its maximum conductivity, which is extraordinary compared to any other 2DEG system. By adopting graphene in 2DEG THz modulators, negligibly low beam attenuation and near unity modulation depth are achievable, offering advantages including RT, broadband, and polarization-independent operation.

ACKNOWLEDGMENT Xing acknowledges the support from National Science Foundation (CAREER award monitored by Pradeep Fulay). Jena and Xing acknowledges the support from National Science Foundation (ECCS-0802125 monitored by Pradeep Fulay) and from the Office of Naval Research (N00014-09-1-0639, Paul Maki). Liu and Xing acknowledge the support from National

Science Foundation (ECCS-1002088 monitored by Andreas Weisshaar). Jena, Xing and Kelly acknowledge the support from the Midwest Institute of Nanoelectronics Discovery (MIND) and the Center for Nanoscience and Technology at the University of Notre Dame. Liu, Jena and Xing also acknowledge the support from the Center of Advanced Diagnostics & Therapeutics at the University of Notre Dame.